\documentclass[lettersize,journal]{IEEEtran}
\usepackage{amsmath,amsfonts}
\usepackage{algorithmic}
\usepackage{array}
\usepackage[caption=false,font=normalsize,labelfont=sf,textfont=sf]{subfig}
\usepackage{textcomp}
\usepackage{stfloats}
\usepackage{url}
\usepackage{verbatim}
\usepackage{graphicx}
\usepackage{bm}
\usepackage{cite}

\def\BibTeX{{\rm B\kern-.05em{\sc i\kern-.025em b}\kern-.08em
    T\kern-.1667em\lower.7ex\hbox{E}\kern-.125emX}}
\usepackage{balance}
\begin{document}
\bstctlcite{BSTcontrol}
\title{Harnessing the Power of AI-Generated Content for Semantic Communication}
\author{Yiru Wang, Wanting Yang, Zehui Xiong,~\IEEEmembership{Senior~Member,~IEEE, }Yuping Zhao, 

Tony Q. S. Quek,~\IEEEmembership{Fellow,~IEEE, }and Zhu Han,~\IEEEmembership{Fellow,~IEEE}
\thanks{
Yiru Wang is with Peking University, China, and also with Singapore University of Technology and Design, Singapore (yiruwang@stu.pku.edu.cn);
Wanting Yang, Zehui Xiong and Tony Q. S. Quek are with the Pillar of Information Systems Technology and Design, Singapore University of Technology and Design, Singapore (e-mail: wanting\_yang@sutd.edu.sg; zehui\_xiong@sutd.edu.sg; tonyquek@sutd.edu.sg);
Yuping Zhao is with Peking University, China (yuping.zhao@pku.edu.cn);
Zhu Han is with the University of Houston, USA (hanzhu22@gmail.com).

}}

\markboth{Journal of \LaTeX\ Class Files,~Vol.~18, No.~9, September~2020}%
{How to Use the IEEEtran \LaTeX \ Templates}

\maketitle

\begin{abstract}
Semantic Communication (SemCom) is envisaged as the next-generation paradigm to address challenges stemming from the conflicts between the increasing volume of transmission data and the scarcity of spectrum resources. However, existing SemCom systems face drawbacks, such as low explainability, modality rigidity, and inadequate reconstruction functionality. Recognizing the transformative capabilities of AI-generated content (AIGC) technologies in content generation, this paper explores a pioneering approach by integrating them into SemCom to address the aforementioned challenges. We employ a three-layer model to illustrate the proposed AIGC-assisted SemCom (AIGC-SCM) architecture, emphasizing its clear deviation from existing SemCom. Grounded in this model, we investigate various AIGC technologies with the potential to augment SemCom's performance. In alignment with SemCom's goal of conveying semantic meanings, we also introduce the new evaluation methods for our AIGC-SCM system. Subsequently, we explore communication scenarios where our proposed AIGC-SCM can realize its potential. For practical implementation, we construct a detailed integration workflow and conduct a case study in a virtual reality image transmission scenario. The results demonstrate our ability to maintain a high degree of alignment between the reconstructed content and the original source information, while substantially minimizing the data volume required for transmission. These findings pave the way for further enhancements in communication efficiency and the improvement of Quality of Service. At last, we present future directions for AIGC-SCM studies.

\end{abstract}

\section{Introduction}
\IEEEPARstart{T} {he} ongoing advancement of artificial intelligence algorithms and the growing support from computational devices have led to the emergence of innovative applications such as autonomous driving, virtual reality (VR), augmented reality (AR) and mixed reality (MR). These applications, in turn, require substantial communication data capacity. At the same time, conventional communication systems are approaching the Shannon limit and the available spectrum resources are becoming increasingly scarce. Therefore, a novel approach to tackle the ever-evolving challenges in communication is greatly needed.

Currently, Semantic Communication (SemCom) has shown its potential as a promising communication paradigm in six-generation (6G), situated at the intersection of artificial intelligence and communication technology. Its primary goal is to convey semantic information from the source to the receiver, focusing on the meaning of the message rather than the precise reception of individual symbols or bits. In contrast to Shannon communication, SemCom adopts the `extract before coding' approach. By delicately devising a pair of semantic codecs, the semantic information concealed within the highly compressed transmitted messages can be reconstructed at the receiver. This approach proves advantageous in scenarios characterized by communication resource constraints and strict privacy preservation requirements.

At present, the outstanding performance of SemCom has been demonstrated in recent literature \cite{wanting}. However, there are still some problems hindering the large-scale implementation of SemCom:
\begin{itemize}
	\item \textbf{Low Explainability:} Traditional SemCom relies on end-to-end models that are hard to interpret, making troubleshooting and manual review both time-intensive and expensive.
	\item \textbf{Modality Rigidity:} The uni-modal designs in current SemCom studies are commonplace, but they tend to be insufficient in scenarios involving multiple modalities or cross-modality interactions. This limitation hinders the adaptability of the existing structure to various communication scenarios.
        \item \textbf{Reconstruction Limitation:} Existing methods fall short in recovering high-quality content when the compression rate is reduced to an extreme level. This limitation hinders the adoption of SemCom, especially in scenarios where communication resources are extremely scarce.
\end{itemize}

Addressing these challenges requires a paradigm shift in SemCom methodologies. In some communication scenarios, the receiver may not require perfect reconstruction of all source information; instead, it prioritizes the accurate presentation of semantic meanings related to the tasks. For example, in vehicular networks, the decisions made by drivers remain unaffected by alterations in the coloration of surrounding vehicles. Similarly, when developing a digital twin for a manufacturing facility, the precise texture or hue of machinery may not hold significance for the simulation and analysis of the plant's operational efficiency. Additionally, within a video monitoring system designed for traffic management, variations in the background environment do not influence the system's capability to identify and track vehicles. Therefore, bandwidth for transmitting less-important information of this nature should be conserved. In certain situations, cross-modality communication is also preferred, exemplified by the choice to transmit direct voice alarms instead of complex radar point clouds to nearby drivers in emergency contexts. In these scenarios, AI-generated content (AIGC) technologies demonstrate their versatility in enhancing communication, due to their numerous advantages, including:
\begin{itemize}
	\item \textbf{High-quality Generation:} AIGC technologies demonstrate remarkable abilities in generating high-quality content in a variety of domains. These capabilities stem from their ability to understand and learn patterns from vast amounts of data during the training process. 
	\item \textbf{Multi-modality Compatibility:} These technologies exhibit remarkable multi-modality capabilities, allowing them to generate and transform content across various data types \cite{comprehensive}, thereby fostering innovation across a wide range of applications.
	\item \textbf{Personalization:} AIGC technologies offer a remarkable personalization ability by harnessing individual preferences and requirements to create tailored content. These technologies can adapt responses and outputs to meet various demands, thus enhancing engagement and satisfaction \cite{gl}. 
\end{itemize}

Through seamless integration of AIGC into SemCom, we can address the aforementioned challenges and endow the AIGC-assisted SemCom (AIGC-SCM) with the following features:
\begin{itemize}
	\item \textbf{Separated Functional Components:} AIGC technologies can be effortlessly integrated with Shannon communication, facilitating the establishment of AIGC-SCM in a plug-and-play fashion. This integration not only boosts interpretability but also optimizes the use of existing infrastructure.
	\item \textbf{Multi-modality Support:} AIGC possesses the capability to generate content in various formats, thereby enhancing SemCom's adaptability to multi-modal tasks. Moreover, by accessing user knowledge bases (KBs)\footnote{A knowledge base (KB) is a centralized repository designed for the organized and efficient management and retrieval of information. Within the SemCom system, the KB encompasses user preferences, network parameters, training datasets, task objectives, and historical communication records.}, it can learn from user attributes and customize content accordingly, ensuring that each SemCom user receives content tailored to their specific needs.
        \item \textbf{Generation-assisted Reconstruction:} Leveraging the generative capability of AIGC, AIGC-SCM demonstrates the ability to reconstruct content with high fidelity and semantic consistency, even in cases of heavily compressed transmitted data. 

\end{itemize}

\begin{figure}
	\centering
	\includegraphics[width=9cm]{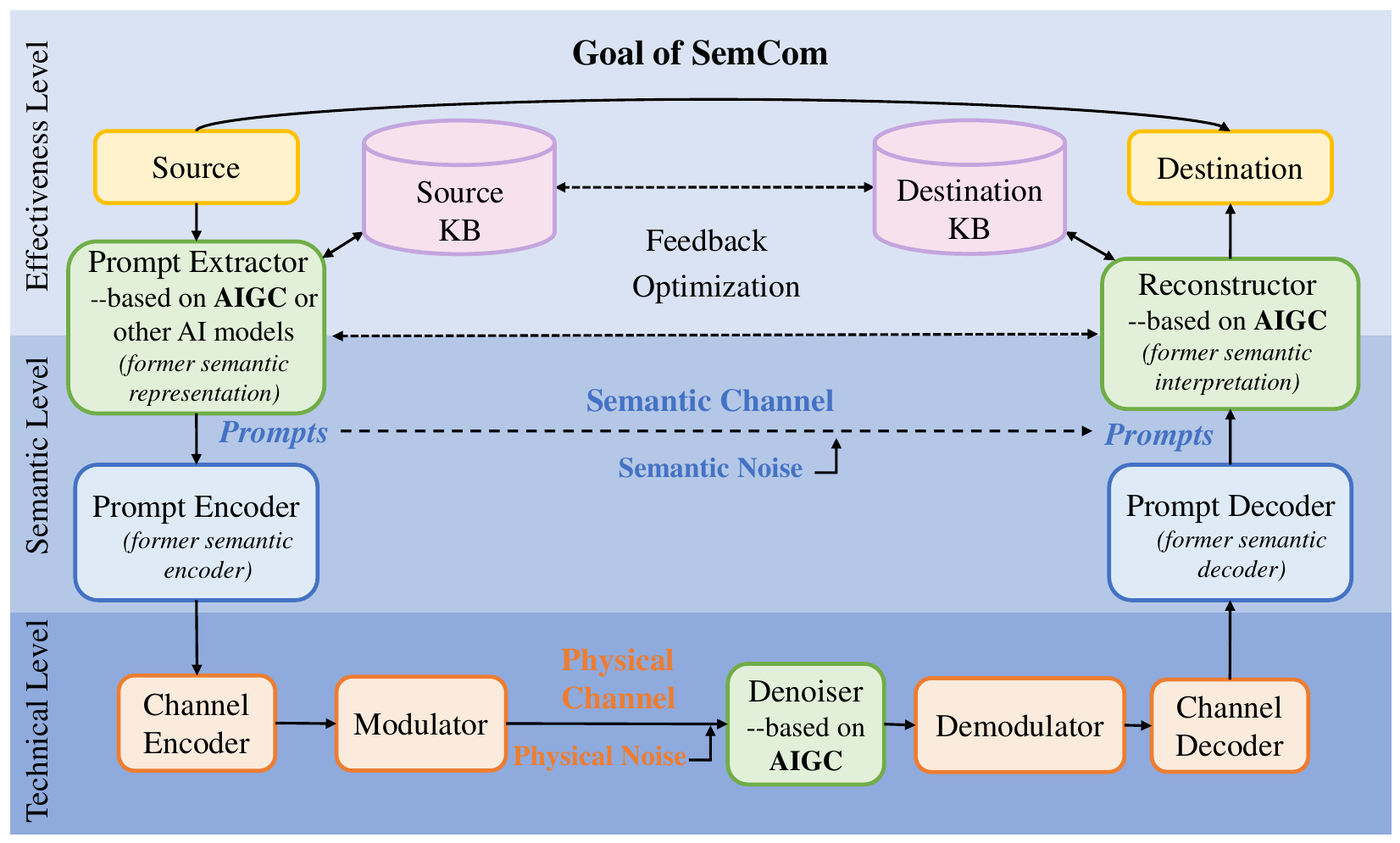}
	\caption{Three-layer model of the proposed AIGC-SCM. Within this framework, semantic features are identified in the form of prompts. Following the traditional source/channel encoding and modulating steps, the prompt is transmitted through the physical channel to the receiver. After undergoing the corresponding demodulating and channel/source decoding procedures, the prompt is recovered to guide generation at the receiver.}
	\label{fig:three-layer}
\end{figure}

While AIGC has demonstrated its versatility in various applications, integrating it into SemCom requires comprehensive investigations. {\em Two challenges} arise when using AIGC to enhance SemCom. Firstly, there is a need to establish a prompt extraction model that aligns with the receiver's tasks, striking a balance between service quality and communication costs. Second, the output of AIGC can be somewhat random and unpredictable. However, we expect to maintain the semantic consistency between the source and the destination in SemCom. Hence, it is crucial to consider strategies for controlling the generation process within AIGC-SCM. Although the authors in \cite{xia} initiate the integration of SemCom and AIGC, the exploration into AIGC's varied roles within SemCom is not exhaustive. Additionally, the impact of diverse prompts and KBs on reconstructing content is not thoroughly examined, pointing towards areas requiring deeper exploration to fully leverage AIGC's capabilities in SemCom contexts.

To fill in the gap, in this paper, we present a unified approach to provide guidance for constructing AIGC-SCM. The contributions of this paper are summarized as follows:
\begin{itemize}
	\item We introduce our proposed AIGC-SCM by rebuilding the three-layer model and highlighting how it differs from the existing structure. To elucidate how AIGC can enhance SemCom, we explore various AIGC technologies and showcase their potential roles as extractors, denoisers, and reconstructors in the SemCom process. Subsequently, we introduce evaluation methods for AIGC-SCM and delve into potential applications.
	\item To address the aforementioned challenges in the practical integration of AIGC into SemCom, we outline step-by-step workflows targeted at different functional aspects. By following these procedures, AIGC-SCM can improve interpretability, achieve multi-modality compatibility, and enhance reconstruction ability.
	\item In our case study, we follow the proposed workflows and exemplify an application of AIGC-SCM in image transmission. We devise specific prompts to cater to diverse application requirements and examine how variations in KBs affect the reconstruction outcomes. Our study illustrates the effectiveness of AIGC-SCM in minimizing bandwidth usage while preserving semantic similarity. 
\end{itemize}

\section{Insights into AIGC-SCM}\label{Insights}


We present a refined three-layer model of AIGC-SCM, as illustrated in Fig. \ref{fig:three-layer}. Different from existing SemCom frameworks, our approach substitutes the semantic representator and semantic encoder with a prompt extractor and prompt encoder at the transmitter. At the receiver, we incorporate an AIGC-based denoiser to mitigate the impact of physical noise. Subsequently, we adapt the semantic decoder and interpreter to function as a prompt decoder and AIGC-assisted reconstructor. In the sequel, we embark on a comprehensive exploration of AIGC, introduce evaluation methods for AIGC-SCM and outline various applications.

\subsection{How AIGC can empower SemCom}\label{why}

\begin{figure*}
	\centering
	\includegraphics[width=18cm]{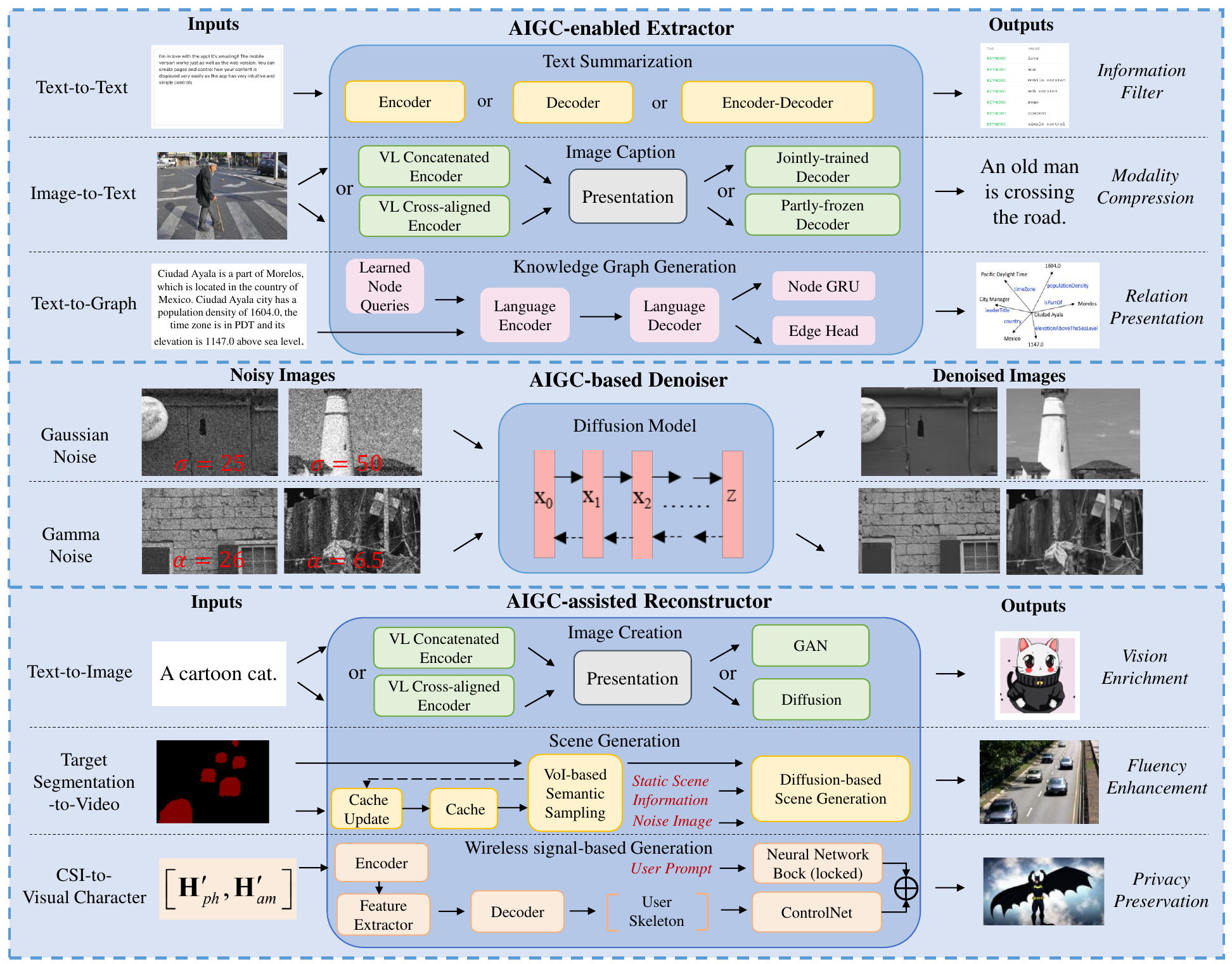}
	\caption{Functions and representative technologies of AIGC for enhancing SemCom. In our proposed AIGC-SCM framework, AIGC can be utilized as the prompt extractor, denoiser, and reconstruction. We highlight several representative AIGC technologies capable of effectively fulfilling these functions.}
	\label{fig:AIGC}
\end{figure*}

In the context of our refined three-layer framework, the incorporation of AIGC fulfills diverse roles. It can operate as the prompt extractor, serve as the denoiser, and play a role as the reconstructor at the receiver. Subsequently, we will elaborate on these functions, accompanied by representative examples illustrated in Fig \ref{fig:AIGC}.

\begin{enumerate}
	\item \textit{For prompt extractor:} The goal of the prompt extractor is to distill semantic prompt from the source data. In scenarios where data is high-dimensional, AIGC can efficiently capture and extract the most informative features. As discussed in \cite{comprehensive}, AIGC possesses the capability to transform the source into an alternative, yet more concise modality, thereby diminishing dimensionality and conserving transmission bandwidth. For example, it can summarize texts from expansive vocabularies, extract descriptive captions from images, and transform lengthy sentences into knowledge graphs.
 
 
	
	\item \textit{For denoiser:} Some AIGC models, such as diffusion models, demonstrate notable capabilities in denoising. By utilizing a progressive denoising approach, they can iteratively enhance content quality. This approach ensures that denoised content remains faithful to the original. This proficiency proves effective across various modal types and noise levels \cite{diffusionnoise}. Therefore, through the application of these models, there is potential to mitigate physical noise across various channel conditions. 
		
	\item \textit{For reconstructor:} In contrast to the extractor, the reconstructor aims to enhance the receiver's quality of experience by generating high-dimensional semantic information based on low-dimensional prompts. In this field, AIGC possesses formidable capabilities and vast applications. For instance, AIGC can also translate text instructions into images \cite{comprehensive}. To address the challenges posed by constrained bandwidth resources in video transmission, a proposed solution is to transmit only the changes in target semantic maps. These changes then serve as guidance for the AIGC-assisted reconstructor in generating video frames \cite{yang2023semantic}. To enhance privacy preservation by minimizing users' exposure to cameras, \cite{CSI} leverages channel state information (CSI) to predict a user's skeleton, which is then employed to instruct the AIGC at the receiver to generate a virtual character.

\end{enumerate}

\subsection{How to evaluate AIGC-SCM performance}\label{Metrics}

\begin{figure*}
	\centering
	\includegraphics[width=18cm]{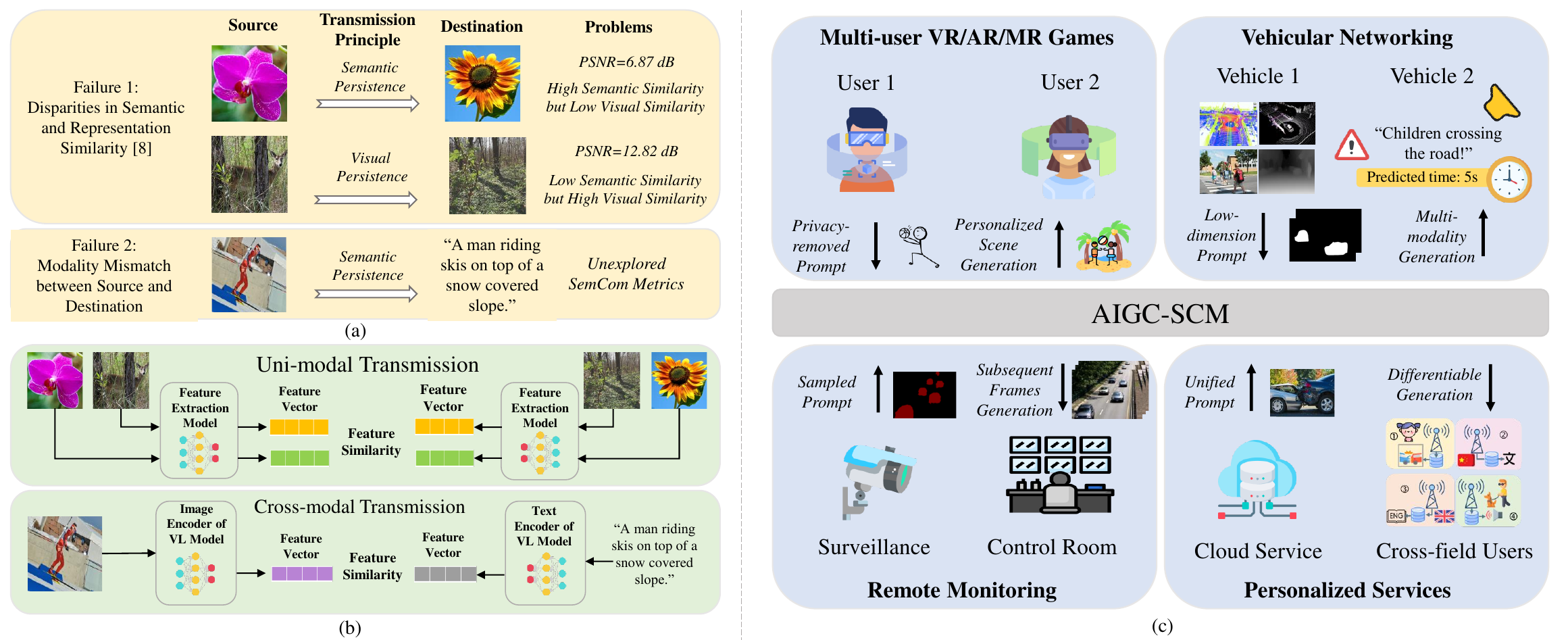}
	\caption{Evaluation methods and applications of AIGC-SCM. Certain conventional metrics may fall short in evaluating AIGC-SCM due to disparities in semantic and representation similarity, along with modality mismatches between the source and destination. To address this, we introduce evaluation methods utilizing feature space vectors for both uni-modal and cross-modal transmission. Additionally, we illustrate how AIGC-SCM can enhance 6G in various application scenarios, including multi-user VR/AR/XR games, vehicular networking, remote monitoring, and personalized services.}
	\label{fig3}
\end{figure*}

As highlighted by \cite{wanting}, unlike metrics such as bit and symbol error rate in Shannon communications, metrics for assessing SemCom should focus on semantic similarity. In this context, conventional metrics may prove inadequate. To illustrate, we present a summary of two scenarios, each accompanied by its corresponding rationale:



\begin{enumerate}
	\item {\em Disparities in Semantic and Representation Similarity \cite{similarity}:} Even when entities share semantic traits, differences in their representation can cause traditional metrics to yield low similarity scores. This happens if the AIGC-assisted reconstructor at the receiver interprets semantic information differently than the transmitter, perhaps through varied sentence expressions or image styles. For example, in image transmission, orchids and sunflowers share high semantic similarity as flowers, but conventional metrics like peak signal-to-noise ratio (PSNR) rate their similarity lower than that between a deer and a forest, due to emphasis on representation consistency.
	\item {\em Modality Mismatch between Source and Destination:} Modality transitions are commonplace in AIGC-SCM as they enable data compression and provide personalized services. However, existing SemCom works have not yet developed corresponding metrics for scenarios where the source and destination exist in different modalities.
\end{enumerate}

In this article, we propose to evaluate AIGC-SCM's performance by comparing the vectors in the feature space between the transmitter and the receiver. This choice is motivated by several advantages: 
\begin{enumerate}
	\item Feature-based similarity often captures the semantic relevance of content more effectively;
	\item By measuring similarity in the feature space, we can mitigate the effects of variations, noise, and modality mismatching;
	\item Features learned from some pre-trained models, like CLIP \cite{CLIP}, a pre-trained vision-language (VL) model that learns to map images and text descriptions into a shared embedding space, are more likely to generalize well to diverse datasets. The encoded high-level patterns are transferable across different tasks and domains. 
\end{enumerate}

In Fig. \ref{fig3}(b), we introduce evaluation methods using the feature space vector for both uni-modal and cross-modal transmission. For uni-modal transmission, we consider image transmission as an example. In this case, the same feature extraction model, such as an image classification model without the final decision layer, can be employed at both the transmitter and receiver to obtain latent feature vectors. A comparison of these vectors can then be calculated using methods like cosine similarity. In the context of cross-modality transmission, we consider image-text transmission as an example. Here, we can choose a pre-trained VL model and input both the image and text into this model. This process allows us to retrieve their feature vectors in a shared embedding space. Finally, an evaluation can be made using standard similarity calculations.

\subsection{How AIGC-SCM can empower 6G}
As depicted in Fig. \ref{fig3}(c), we illustrate how AIGC-SCM can enhance 6G in four typical application scenarios, described as follows:

\begin{enumerate}
	\item \textit{Multi-user VR/AR/MR Games:} Several challenges have emerged in achieving the large-scale implementation of VR/AR/MR gaming, including restricted communication resources, privacy concerns, and monotonous content. A promising solution to address these issues is AIGC-SCM, as it can represent virtual scenes within the prompt and generate diverse and personalized content for different users.
	\item \textit{Vehicular Networking:} Extensive data exchanges from numerous sensors in vehicular networks exert a considerable burden on transmission systems. To mitigate this, AIGC-SCM can be employed to distill and transmit location-related information. At the receiver end, it can enable the intuitive recognition and display of traffic congestion and unusual incidents, providing essential information for drivers' decision-making.
	\item \textit{Remote Monitoring:} In typical and foreseeable remote monitoring scenarios, AIGC-SCM exhibits the capability to reconstruct the current frame based on received prompts and predict subsequent frames conditioned on it. This capacity offers a solution to reduce the volume of communication data.
	\item \textit{Personalized Services:} AIGC-SCM is equipped with a remarkable personalization ability by harnessing individual preferences and behavior data to adapt responses and outputs to cater to specific receivers, enhancing engagement and satisfaction. 
\end{enumerate}

\section{AIGC-SCM Framework Implementation}\label{Framework}

\begin{figure*}
	\centering
	\includegraphics[width=18cm]{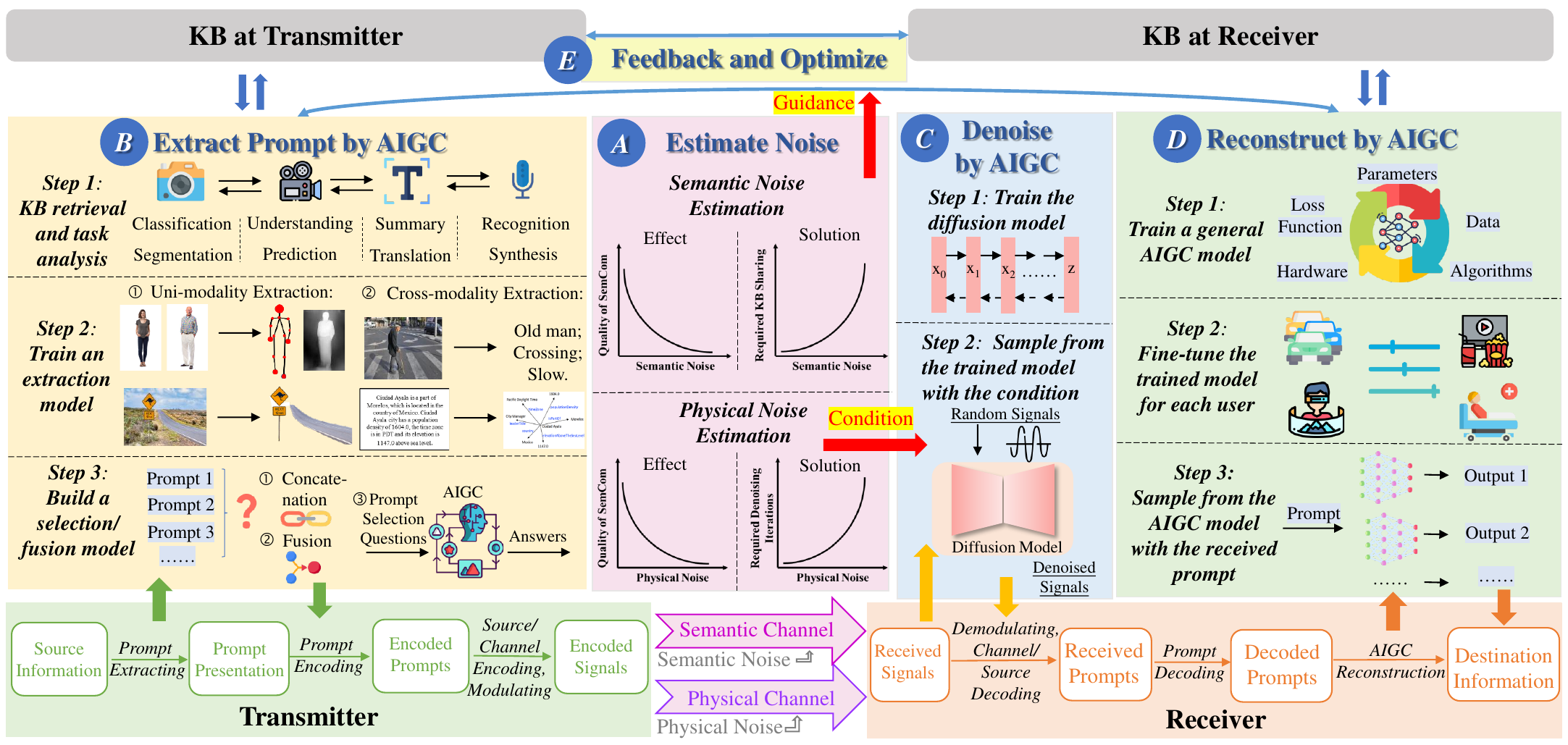}
	\caption{Workflow of implementing AIGC-SCM. The AIGC-SCM implementation workflow involves a systematic process, beginning with the estimation of noise as a preparatory step. Subsequently, AIGC is employed in three pivotal stages: extracting prompts, denoising received signals, and reconstructing contents. The final step entails gathering feedback and optimizing the entire system.}
	\label{fig:picture001}
\end{figure*}

While the integration of AIGC into the current framework poses potential benefits for SemCom, it is not straightforward. We need to properly control the AIGC generation and ensure that AIGC-SCM meets the receiver's requirements. In this section, we discuss the specific procedures that need to be implemented to ensure the smooth functioning of the entire system.

\subsection{Estimate Noise}\label{Est} 
Both physical noise and semantic noise can impact the reconstruction process and reduce the quality of SemCom. Nonetheless, by accurately estimating their levels, we can optimize the system design to mitigate their effects accordingly. The estimation of physical noise can serve as a condition for the AIGC-assisted denoiser, influencing the determination of denoising iterations. Semantic noise estimation serves as guidance for knowledge sharing. In instances where semantic noise is high, enabling the sharing of data, models, and domain-specific terminology is advisable before engaging in communication.



\subsection{Extract Prompt by AIGC}\label{How1}

In AIGC-SCM, the prompts are extracted at the transmitter to distill the semantic meanings from the source.  Given that both modality and the quantity of prompts impact the system's performance, a decision on the extraction model is crucial. To address this, we provide a general procedure for designing the extraction model and generating effective prompts:

\begin{itemize}
	\item \textbf{Step 1: KB retrieval and task analysis.} In SemCom, the KBs store information about each user's datasets, preferences, and tasks. Accessing KBs enables the identification of the specific tasks for the AIGC model, facilitating the development of suitable extraction models.
	\item \textbf{Step 2: Train an extraction model.} Following task analysis, we can tailor the design of our prompt extraction models accordingly. Traditional deep learning models, including CNNs, RNNs, and Transformers, can be intentionally designed and trained to extract uni-modality semantic features. Furthermore, AIGC models can accomplish modality transformation, providing a method to extract cross-modality prompts and hence reduce data transmission volume, as previously discussed in Section \ref{why}.
	\item \textbf{Step 3: Build a selection/fusion model.} In situations where communication resources are constrained, the determination of final prompts from an extensive set of acquired features becomes crucial. One approach involves building reinforcement learning models to select and concatenate the most critical features. Another option is to leverage deep learning models for feature fusion, creating a more compressed prompt. Additionally, certain AIGC models, functioning as decision-makers, can be utilized to assist in prompt selection. 
\end{itemize}

\subsection{Denoise by AIGC} 

When the signal arrives at the receiver, it has been distorted by the physical channel. As we discuss in  Section \ref{why}, some AIGC models can be utilized for denoising communication signals. We take the diffusion model for example and discuss the following steps for denoising:
\begin{itemize}
	\item \textbf{Step 1: Train the diffusion model.} During diffusion model training, there are two key processes: the forward process and the backward process. In the forward process, the diffusion model gradually adds noise to an input data sample. Subsequently, the diffusion model proceeds with the backward process and reverses the degradation. Starting with the most noisy version of the data, the model's objective is to denoise the data iteratively, attempting to recover the original, noise-free input data. In light of the above, the model learns to remove noise and restore data to its original quality.  
	\item \textbf{Step 2: Sample from the trained diffusion model with the condition.} After the training, the sampling procedure leverages the trained network to generate a denoised sample. With the condition of the estimated physical noise, we can reduce the discrepancy between the modified and true conditional distributions for each sampling step, thus making the ultimately recovered signal much closer to the original transmitted signal.
\end{itemize}

\subsection{Reconstruct by AIGC} 

Following the decoding process, the prompt is recovered, serving as a condition for the implementation of an AIGC-based reconstruction. In the single-user scenario, we can directly train a reconstruction model that satisfies the user's requirement. In the multi-user scenario, the sender can likewise transmit only one prompt to enhance the communication efficiency, because the receiver can employ the AIGC model to generate personal content in a variety of presentation formats based on the same semantic information. To ensure that the trained model is adaptable to various user scenarios, we employ an efficient three-step strategy, outlined as follows:

\begin{itemize}
	\item \textbf{Step 1: Train a general AIGC model.} First, we develop a general model, like a stable diffusion model \cite{stable} for text-to-image transmission. Before training, we should collect the dataset for this model, which is typically extensive and diverse within the modalities and tasks defined. It can enable the basic model to effectively execute different tasks and generalize across various scenarios. 
	\item \textbf{Step 2: Fine-tune the trained model.} In a multi-user scenario, receivers often have diverse domains and preferences. To enable the AIGC model to produce personalized content, we need to fine-tune the general model based on personal databases. For instance, in the text-to-image transmission scenario, we fine-tune our trained diffusion model to facilitate image reconstruction in various styles. Even in a single-user scenario, this step proves beneficial when the user's preferences change.
	\item \textbf{Step 3: Sample from the AIGC model with the received prompt.} After training the AIGC generation model, we integrate it into the AIGC-SCM framework. When the decoded prompt is input into the AIGC models of each user, they have the capability to generate more tailored semantic interpretations.
\end{itemize}

\subsection{Feedback and Optimize} 

The described method initiates the AIGC-SCM model, yet practical use demands ongoing feedback-driven refinement, including KB sharing and fine-tuning between transmitter and receiver. This feedback is vital for the AIGC-SCM's continuous enhancement and its adaptation to users' changing needs and the dynamic communication environment.

\section{Case Study: AIGC-SCM for Image Transmission}\label{Case}

\begin{figure}
	\centering
	\includegraphics[width=8cm]{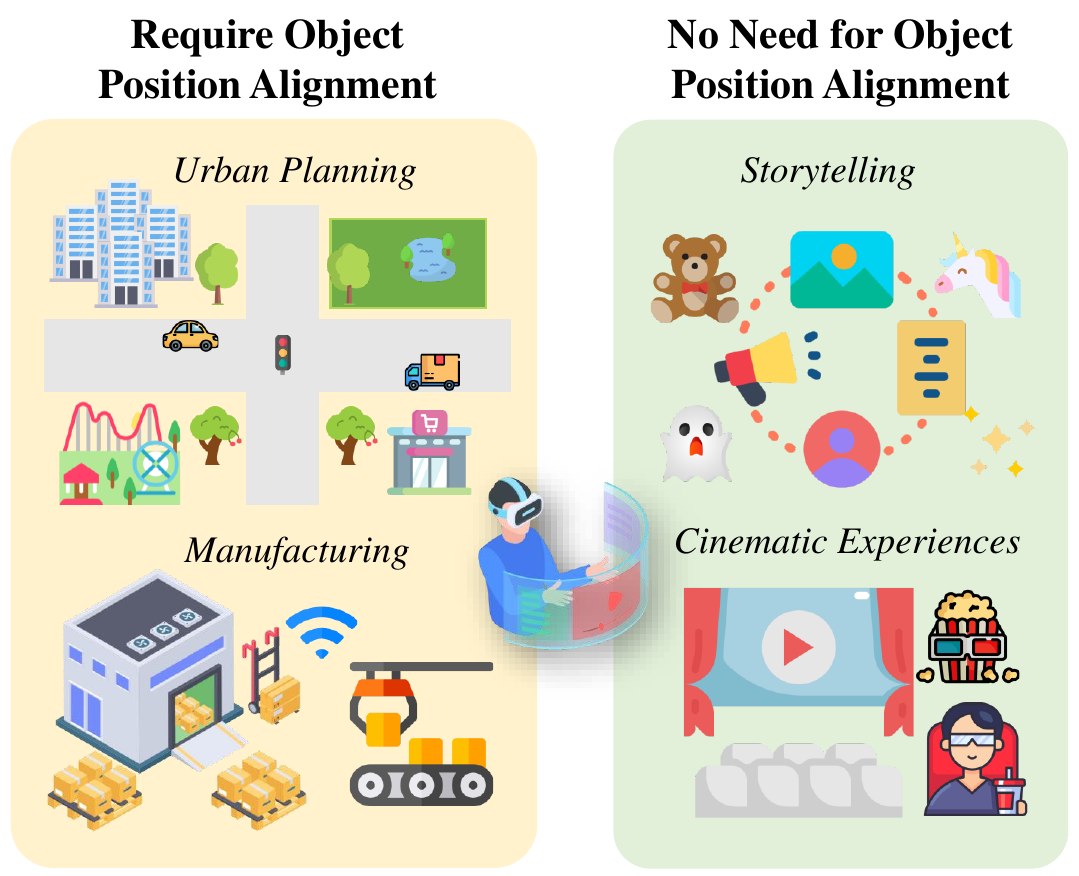}
	\caption{VR image transmission scenarios categorized based on object position alignment requirements. The left two applications exemplify VR mapping, demanding high accuracy in object position. In contrast, the misalignment of object position between the transmitter and the receiver does not impact the experience of the right two applications. }
	\label{fig5}
\end{figure}

Communication plays a pivotal role in VR applications, serving a multitude of purposes. It enables the facilitation of multiplayer experiences, the establishment of collaborative workspaces, and the enhancement of immersive experiences. However, the substantial data volume and increased latency associated with traditional VR image transmission lead to a diminished VR experience. To address this, we propose to use AIGC-SCM for data transmission, which can efficiently extract and reconstruct semantic information and reduce the required communication resources. As shown in Fig. \ref{fig5}, to differentiate between various VR application requirements and design extraction models accordingly, we classify image transmission scenarios based on whether they require the alignment of object positions between the transmitter and receiver images.

\subsection{AIGC-SCM Schemes and KB Differentiation}

In our case study, we examine a VR image transmission scenario where Transmitter A intends to multicast its scene to two distinct recipients. To assess the impact of KBs on the reconstruction process, we assume that Receiver B possesses only general knowledge, while Receiver C has additional historical information about the objects created within Transmitter A's scene.

To accommodate scenarios with object position alignment requirements (A) or without object position alignment requirements (I), and to tailor for users with historical (H) knowledge or without historical (N) knowledge, we implement four distinct AIGC-SCM schemes: NI, NA, HI, and HA. 

For the prompt extraction processes, we employ the YOLO object detection model \cite{YOLO} to extract labels and bounding boxes in NA and HA schemes. In NI and HI schemes, we utilize the image caption ExpansionNet v2\footnote{https://github.com/jchenghu/ExpansionNet\_v2} model to extract text descriptions. This extracted text, along with unique identifiers\footnote{In practice, every transmission requires a unique identifier to accurately locate the object mentioned in the prompt within Receiver C's historical data. For Receiver B, this identifier can be considered semantic noise and thus disregarded. For the sake of clarity and conciseness, the transmission of these short identifiers is not taken into account in the visual representation of our results or the numerical comparisons presented in this paper.}, serves as the prompt for transmission. 

For the image reconstruction processes, we employ the stable diffusion v2\footnote{https://huggingface.co/CompVis/stable-diffusion-v1-2} model for NI scheme and the Dreambooth model \cite{Dreambooth} for HI scheme, and the GLIGEN \cite{gl} model for both NA and HA schemes.

 The LAION \cite{laion} and COCO2014 \cite{COCO} datasets are considered as general knowledge, which is stored in both receivers' KBs. The LAION \cite{laion} dataset is regarded as the general knowledge due to its role in training the stable diffusion model \cite{stable}, which is subsequently utilized as the pre-trained model for both receivers. Encompassing around 400 million image-text pairs, this dataset encapsulates a broad array of subjects, themes, and visual styles. Another dataset serving as general knowledge is the COCO2014 \cite{COCO} dataset, which features over 200,000 labeled images across 80 object categories and is utilized in the training of the GLIGEN \cite{gl} model for the NA and HA schemes.

Meanwhile, the Dreambooth dataset \cite{Dreambooth}  includes 30 subjects of 15 different classes and contains 4-6 images per subject, which are captured in different conditions, environments and under different angles. In our study, the images in this dataset represent specific historical information shared by Transmitter A with Receiver C, stored in Receiver C's KB beyond general knowledge. Nonetheless, the application of this dataset diverges between the HI and HA schemes. Within the HI framework, the Dreambooth \cite{Dreambooth} model is adopted, which is fine-tuned on a pre-trained stable diffusion model \cite{stable} using this dataset. When the prompt is decoded, the Dreambooth \cite{Dreambooth} model can utilize its prior knowledge of the subject and generate class-specific instances according to the prompt's context.  
Conversely, in the HA configuration, upon prompt decoding at the receiver, a pertinent historical image from the Dreambooth \cite{Dreambooth} dataset is fetched and integrated into the embedding layers of the GLIGEN \cite{gl} model to guide the image reconstruction.

\begin{figure*}
	\centering
	\includegraphics[width=18cm]{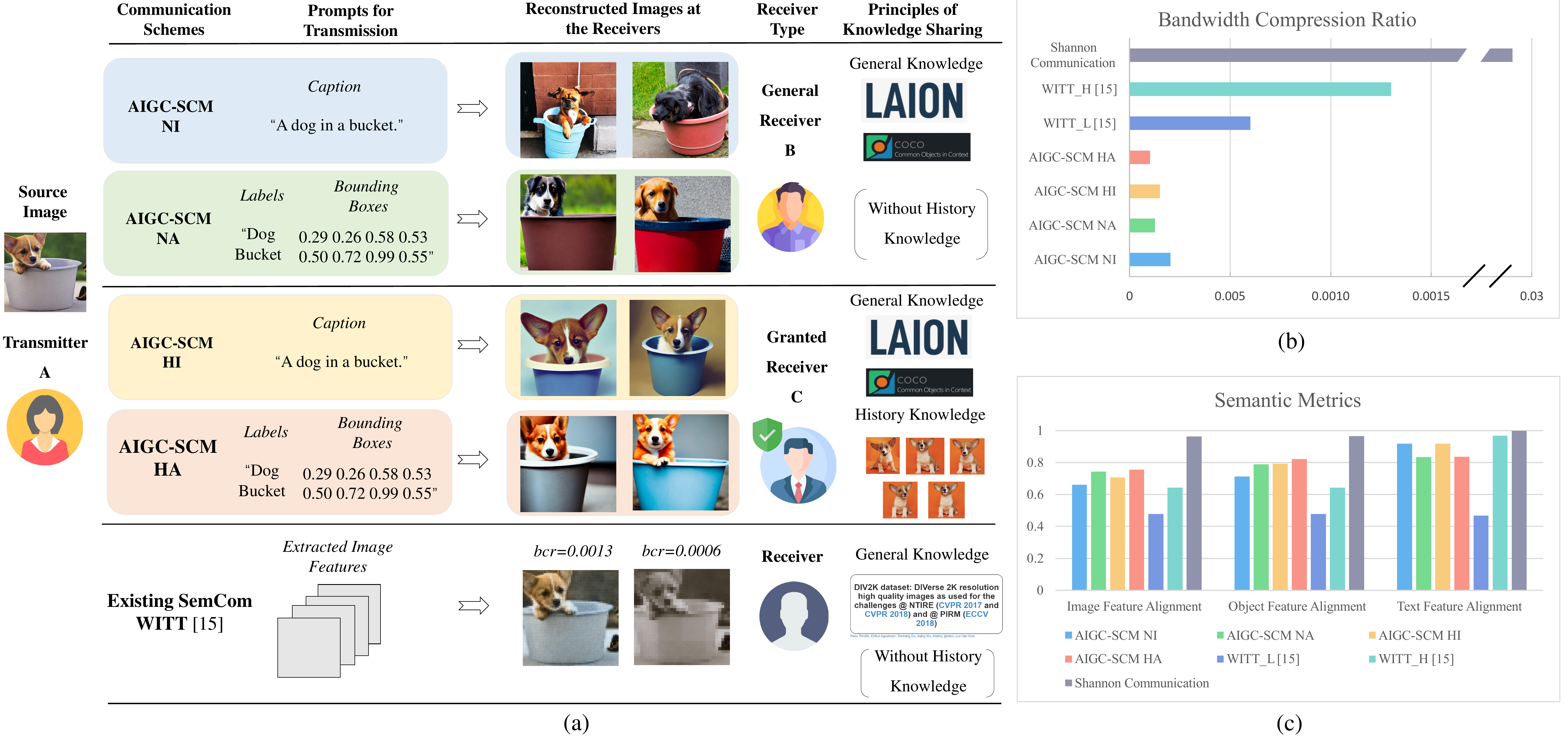}
	\caption{Performance comparison of different communication schemes. We compare our proposed AIGC-based SemCom with two benchmarks: 1) Existing AI-based SemCom scheme, where we choose WITT \cite{witt} model with a higher (\_H) and lower (\_L) bandwidth compression ratio. It utilizes the Swin Transformer to manage long-range image dependencies, incorporates a spatial modulation module for adaptability to fluctuating wireless channel conditions, and employs a hierarchical design to streamline high-resolution image processing; 2) Shannon communication, where we use JPEG for direct image compression, ½ LDPC for channel coding, and 16QAM for modulation. All of these schemes undergo testing under the condition of the 10 dB SNR. To enhance the AIGC-SCM's interpretability and ensure its seamless integration with existing communication architectures, the channel-related modules employed in our AIGC-SCM configurations are aligned with those of the Shannon communication scheme.}
	\label{fig6}
\end{figure*}

\subsection{Visual and Numerical Results}

In Fig. \ref{fig6}(a), we present a visual performance comparison between the proposed AIGC-SCM schemes and the existing SemCom scheme WITT \cite{witt}. It is evident that all the schemes successfully preserve a certain level of semantic similarity when compared to the source image, as they all depict images of a dog in a bucket. However, by transmitting labels and bounding boxes, we gain better control over the object positions. Since AIGC-SCM NA and NI can only rely on general KBs, while AIGC-SCM HA and HI can condition on historical knowledge, it becomes apparent that the dogs depicted in schemes AIGC-SCM HA and  HI exhibit a closer resemblance to the source image. In contrast, the reconstructed images generated by the WITT \cite{witt} scheme tend to appear blurry and pale due to its sensitivity to compression, distinguishing its images from the clarity achieved by our proposed schemes under similar compression rates.

In Fig. \ref{fig6}(b), we conduct a comparison involving our four proposed AIGC-SCM schemes, the WITT \cite{witt} scheme, and the Shannon communication scheme. This comparison is based on the average bandwidth compression ratio ($k/n$), which is defined as the ratio of the average number of transmission symbols to the original information bits. The primary difference between our proposed AIGC-SCM and WITT \cite{witt} lies in the fact that the latter persists in utilizing the features of the complete original figure for transmission, while we opt to eliminate extraneous details in the image through image-to-text modality transformation. Additionally, thanks to the use of the conditional generation mechanism and the high-quality generation capabilities of AIGC, this approach can significantly reduce the amount of data transmitted and still reconstruct content with high fidelity and semantic accuracy.

Finally, we assess the similarity performance between the source and destination images. As discussed in Section \ref{Metrics}, some traditional metrics like PSNR are not suitable for evaluating AIGC-SCM. Hence, we extract three types of feature vectors by feeding the data from the source and the destination to CLIP \cite{CLIP} and calculate their cosine similarity for each type. The corresponding metrics and evaluation process can be summarized as follows:
\begin{enumerate}
	\item {\em Image Feature Alignment:} It is measured by calculating the cosine similarity of the whole image features of the source and destination;
	\item {\em Object Feature Alignment:} This metric entails using an object detector to crop the source and destination images and distill only object-related feature vectors. Subsequently, the cosine similarity of these object-related feature vectors is calculated;
	\item {\em Text Feature Alignment:} Given the value of image captions in enhancing image content comprehension, we also gauge the correspondence between the feature vectors of the generated captions of the source and destination.
\end{enumerate}

In Fig. \ref{fig6}(c), we demonstrate the performance of different communication schemes measured using the abovementioned metrics. The highest similarity performance of Shannon communication can be attributed to its utilization of a significantly larger bandwidth resource. Our proposed AIGC-SCM schemes, consisting of four different approaches, outperform the WITT \cite{witt} models in terms of image feature and object feature alignment metrics. This superiority is attributed to AIGC's ability to reconstruct high-quality images with minimal information transmission. Even on the text feature alignment metric, our approach outperforms WITT \cite{witt} when the bandwidth compression rate is low. Schemes that leverage historical data attain elevated alignment metrics, attributed to the model's acquisition of object texture knowledge, enabling more precise reconstructions upon receiving pertinent prompts. Aside from the WITT\_L scheme, text feature alignment scores of other methods significantly outperform traditional semantic alignment metrics. This discrepancy arises because the captioning process distills image content into text, emphasizing salient features as interpreted by the model, yet this abstraction avoids delving into the local details and nuances of the original image. Consequently, images regenerated from captions align with the broad description but miss the precise visual specifics of the originals. Overall, the results illustrate that AIGC-SCM successfully preserves a high degree of semantic similarity while significantly reducing communication bandwidth consumption.

\section{Future Directions}\label{Future}
\subsection{Prompt Selection based on Semantic Noise Estimation}
To mitigate the impact of semantic noise on the quality of SemCom, we can devise prompt selection methods based on the estimated semantic noise. In instances where the noise level is high, we can increase the number of prompts for transmission accordingly, such as transmitting both text description and image features. This approach is analogous to enhancing error-correction ability by expanding code length in traditional channel coding.


\subsection{KB Design in AIGC-SCM}
The contents in the receiver's KB can guide the AIGC-assisted reconstruction process. To conserve transmission bandwidth, historical or less crucial information can be stored in the receiver's KB during the knowledge-sharing process. For instance, in video transmission, storing the object's appearance and background in the first frame enables the prompt transmission of only the changes in movement. Ultimately, the complete scene can be reconstructed by conditioning on both the stored knowledge and the received prompts.

\subsection{Minimizing Latency in AIGC-SCM}
In environments where immediacy is critical, it is imperative to strive for the reduction of the overall system latency, for which various strategies can be employed. For example, model optimization can streamline both the extraction and generation processes, thereby reducing processing times to the millisecond scale \cite{YOLO,comprehensive}. Additionally, the implementation of predictive prefetching can enable the anticipation of future requests, facilitating their rapid resolution and further improving system responsiveness.

\section{CONCLUSION}\label{Conclusion}

This article proposed to use the AIGC-SCM to address challenges in communication systems. We introduced an AIGC-SCM architecture based on a three-layer model and explored some AIGC technologies that can enhance SemCom. Our investigation indicated that thanks to its multi-modality compatibility and high-quality generation capabilities, AIGC-SCM shows promise in enhancing communication across multi-user VR/AR/XR games, vehicular networking, remote monitoring, and the provision of personalized services. We showcased the effectiveness of AIGC-SCM through a case study focused on VR image transmission. The findings from this study illuminated AIGC-SCM's capacity to substantially lower bandwidth requirements while maintaining a high level of semantic similarity to the original information. This not only demonstrated the model's efficiency but also its potential in revolutionizing the way we approach semantic communication in various high-demand applications.

\bibliographystyle{IEEEtran}
\bibliography{IEEEabrv,paper}

\end{document}